\documentclass[a4paper,11pt]{article}
\usepackage{pos}
\usepackage{natbib}
\graphicspath{{figs/}}
\newcommand{\EQ}[1]{Eq.~\eqref{#1}}
\newcommand{\FIG}[1]{Fig.~\ref{fig:#1}}
\newcommand{\SEC}[1]{Sec.~\ref{sec:#1}}
\newcommand{\MSbar}{\overline{\text{MS}}}
\newcommand{\g}{\gamma}
\renewcommand{\a}{\alpha}
\newcommand{\GeV}{\text{ GeV}}
\renewcommand{\O}{\mathcal{O}}

\definecolor{dkgreen}{rgb}{0.0,0.4,0.0}
\newcommand{\green}[1]{{\color{dkgreen} #1}}

\title{$m_c$ (and $m_b$) from lattice QCD}

\author*{Andrew T.\ Lytle}

\affiliation{
Department of Physics, 
University of Illinois at Urbana-Champaign,\\
Urbana, Illinois, 61801, USA}

\emailAdd{atlytle@illinois.edu}

\abstract{
Quark mass determinations based on lattice QCD simulations have continued
to make strides in recent years. Here I review that progress with a focus
on developments computing the charm (and bottom) quark masses since the
2015 edition of CHARM.
These advances have resulted in 
groups now
quoting (sub-)percent-level precision for these quantities, and, importantly, using a variety of techniques subject to
differing systematic uncertainties. 
Improvements to quantify the effects of QED are also now being
included.
I will highlight three of the
strategies being used to determine $m_c$ at this level of precision.
}

\FullConference{%
  *** 10th International Workshop on Charm Physics (CHARM2020), ***\\
  *** 31 May - 4 June, 2021 ***\\
  *** Mexico City, Mexico - Online ***
}

\begin{document}
\maketitle

\section{Introduction}
When preparing this talk, I returned to materials I had prepared
from the 2015 edition of the CHARM conference. It was natural to frame the presentation in terms of
the progress that has been made since that iteration, and I was pleasantly
surprised just how much progress there has been on the topic.
This progress has been partially incremental, in that precision for various
calculations continues to improve as time goes on due to increases
in computing power, refinement of
techniques, more ensembles with finer lattice spacings, etc.\ --
but importantly the number
of techniques being used to achieve quark masses at a high level of precision
has increased, and these results are on the whole in agreement with one another
despite the very different approaches (and corresponding systematic 
sources of uncertainty). This gives
confidence in the robustness of these results and in the reliability of the error estimates, and taken together
represents a significant achievement of the lattice community as a whole.

The goal of this talk then is to summarize the recent results in the field,
paying attention to the various methodologies being used.
In this talk I will focus primarily on three different methods.
(I will also rely on FLAG averages, and specifically highlight new results
available since the last FLAG review~\cite{FlavourLatticeAveragingGroup:2019iem}.) These are the methods of current-current
correlators (which was the focus of the Charm 2015 review~\cite{Lytle:2015oja}),
RI/(S)MOM intermediate schemes, and MRS masses.
\SEC{Intro} will briefly review the definition of quark mass parameters
and how they are determined from lattice QCD simulations.
\SEC{charm} comprises the bulk of the article focused on charm mass, 
and goes into some detail on the methods used by different groups.
\SEC{bottom} provides a brief update on the state-of-the-art in mass
determinations of the bottom quark.
\SEC{Summary} summarizes the results and discusses the future of these calculations.

\section{Quark mass and LQCD} \label{sec:Intro}
Quark masses are fundamental parameters 
that arise in the Standard Model 
from interaction with the Higgs field, 
and the precision determinations of heavy quark masses are 
needed to stringently test Standard Model predictions 
for Higgs-fermion couplings~\cite{Lepage:2014fla}. 
Because of confinement in QCD, quark masses
cannot be measured directly but must be related to physical observables, such as hadron masses, via theoretical tools.
Quark mass parameters are scheme and scale dependent quantities,
and are typically quoted in the $\MSbar$ scheme. In this document
I will try to make explicit this dependence, for the charm and
bottom results are typically quoted at the scale of the
quark itself, i.e. $m_c^{\MSbar}(m_c)$, $m_b^{\MSbar}(m_b)$, though
sometimes other choices are made such as $m^{\MSbar}_c(3 \GeV)$.
The quark mass definitions also depend on the number of
quarks in the sea $n_f$, and I've also tried to make note of this where applicable.

In lattice QCD simulations, quark masses are dimensionless 
input parameters
specified in terms of the lattice spacing $a$, i.e.\ 
$am_{ud,0}$, $am_{s,0}$, $am_{c,0}$. The subscript `0' on these
quantities is to indicate they are bare input quantities, and
depend on the details of the regularization being employed. In
particular they have no well-defined physical meaning, but must
be related to renormalized quark masses defined in the continuum.
(In contrast quark mass \emph{ratios} are physically meaningful,
and can be used for example to obtain $m_b$ from $m_c$
and a determination of $m_b/m_c$.)
In simulation each input quark mass is tuned to reproduce a physical
quantity, for example $am_{ud,0}$ may be tuned to reproduce the
pion mass, while $am_{c,0}$ may be tuned to reproduce the $J/\psi$
mass. After these tunings, and a measurement of the lattice spacing,
the remaining observables are predictions of the theory of QCD.
The ability to dial the quark input parameters and measure the
resulting change in physical quantities is one of the reasons lattice 
simulations are well-suited to precision mass determinations.
In particular we will see that being able to map out 
physical dependences
on fictitious heavy quark masses $am_c < am_h < am_b$ 
is useful. How to make the connection between
bare quantities in simulation, and the continuum renormalized
quantities, will be explored in the subsequent sections.

\section{Charm quark mass} \label{sec:charm}
\begin{figure}
    \centering
    \includegraphics[width=0.6\textwidth]{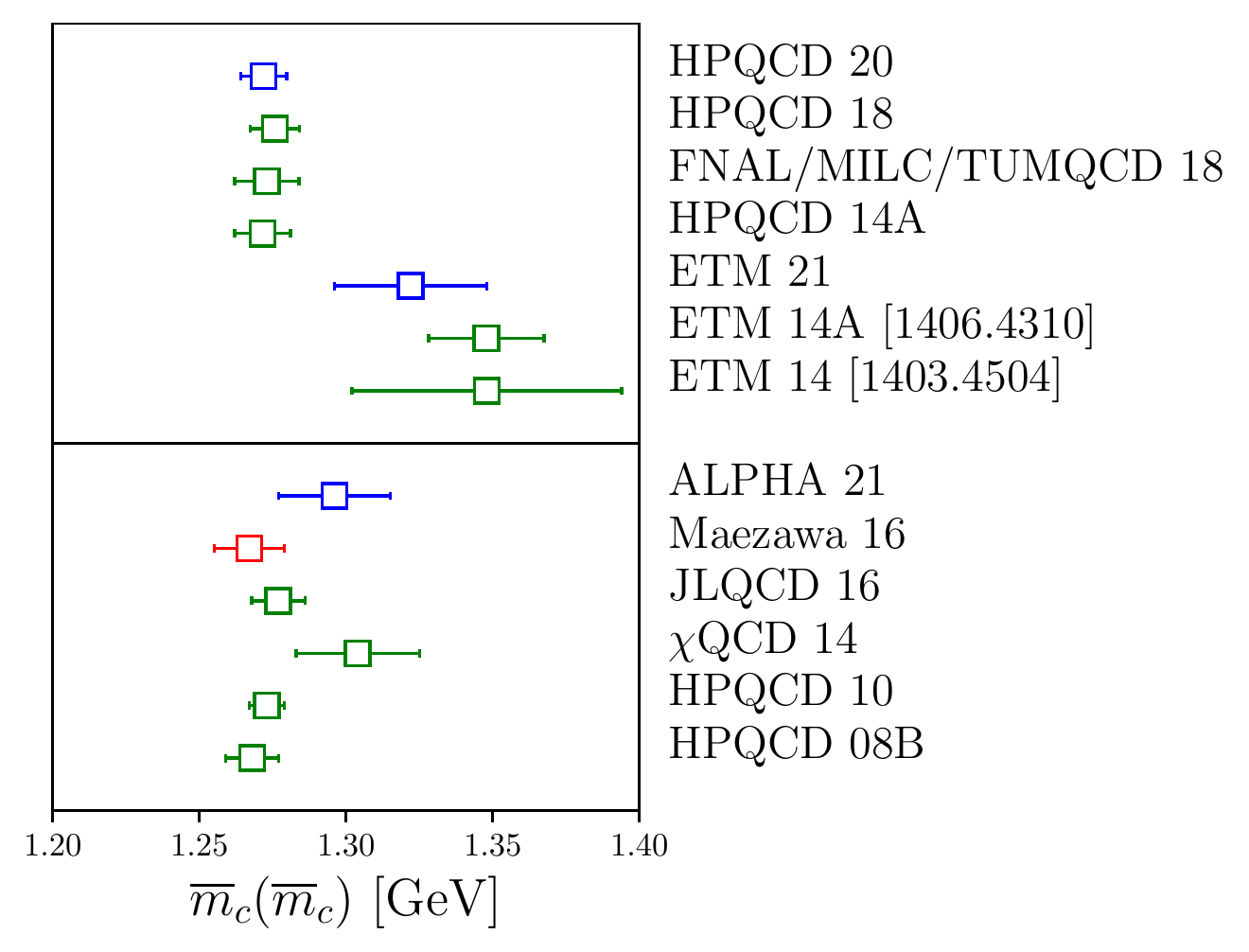}
    \caption{Summary plot of charm mass determinations from the lattice,
    adapted from 2019 FLAG review. The top (bottom) panel gives results
    based on gauge configurations with $n_f=4$ ($n_f=3$) quark flavors
    in the sea. New entries published after the review
    are labelled in blue while the other entries maintain the green/red
    color designation given by FLAG.}
    \label{fig:mclatt21}
\end{figure}

Since 2015 the number of lattice charm mass determinations has more than
doubled from 6 to 13, including three since the 2019 FLAG review.
The situation is summarized in \FIG{mclatt21}, where I have updated the results
of the FLAG review to reflect these new entries. For the purposes of
the talk I would like to especially focus on the four results at the top
of the figure, which all separately claim percent-level uncertainty,
and remarkably are arrived at through three completely different techniques.
I would also like to highlight the new results
based on small volume step-scaling 
techniques~\cite{Heitger:2021apz,Campos:2018ahf}
from ALPHA and a new calculation from ETM 
using RI/MOM renormalization~\cite{Alexandrou:2021gqw}
(a variant
of the type of calculation described in more detail in \SEC{ri/smom}).
These are labelled ALPHA 21 and ETM 21 in \FIG{mclatt21}, respectively.

\subsection{Current-current correlator method} \label{sec:jj-charm}
The lattice current-current correlator method was pioneered
in~\cite{HPQCD:2008kxl,McNeile:2010ji,Chakraborty:2014aca}, and in essence
consists of comparing time-moments of Euclidean current-current
correlators computed on
the lattice with the predictions of high-order perturbation theory.
The correlators (and moments) needed are straightforward to compute on
the lattice. The Euclidean-time twopoint function is given by
\begin{equation}
G(t) = a^6 (a m_{0h})^2 
\sum_{\mathbf{x}} \langle J_5(t, \mathbf{x}) J_5(0, 0) \rangle \, .
\label{eq:G(t)}
\end{equation}
where $J_5 \equiv \bar{\psi}_h \g_5 \psi_h$ and $a m_{0h}$ is the
bare quark mass parameter in lattice units. The quantity in brackets
is simply the pseudoscalar heavyonium $h \bar{h}$ two-point correlator,
from which the particle mass and decay constant would be extracted in
a ``typical'' lattice calculation.
Here $am_{0h}$ is an input mass that can be chosen
to correspond to the charm mass, but
as will be seen it is also convenient to allow this
mass to vary from charm to bottom.
In formalisms with sufficient chiral symmetry,
the current $J_5$ is absolutely normalized, so that no auxiliary calculation
of matching factors is needed. The combination in~\EQ{eq:G(t)} is additionally
UV finite, so that the lattice computed quantity is equal to its continuum
counterpart up to discretization artifacts,
\begin{equation}
G(t)_{\text{cont}} = G(t)_{\text{latt}} + \O(a^2) \qquad (t \neq 0) \, .
\end{equation}
Using this two point calculation the time moments are constructed,
\begin{equation}
G_{n,\text{latt}} = \sum_{t=0}^{T} (t / a)^{n} \, G(t)_{\text{latt}} \,,
\end{equation}
which for a given lattice ensemble and input mass $am_{h0}$ give a set
of pure numbers indexed by $n$. These time moments have also been computed
in continuum perturbation theory to
$\text{N}^3$LO~\cite{Chetyrkin:2006xg,Boughezal:2006px,Maier:2009fz},
and they are sensitive to
the renormalized quark masses.
For $n \geq 4$,
\begin{equation} \label{G_n,pert}
G_{n,\text{pert}} = 
\frac{g_n(\a_{\MSbar}, \mu)}{(am_{h}(\mu))^{n-4}} \, ,
\end{equation}
where $m_h(\mu)$ is the $\MSbar$ quark mass at the scale $\mu$.
The goal then is to find the values for $\a_{\MSbar}(\mu)$ and
$m_h(\mu)$ so that the $G_{n, \text{pert}}$ agree with the lattice
data $G_{n, \text{latt}}$, up to discretization artifacts.

In practice HPQCD carries out the analysis in terms of reduced moments
which are simply related to the time-moments as
\begin{align}
R_4 &= G_4 / G_4^{(0)} \\
R_n &= \frac{1}{m_{0c}} (G_n / G_n^{(0)})^{1/(n-4)} \quad (n \geq 6) \,.
\label{R_n,latt}
\end{align}
where $G_n^{(0)}$ are the tree-level results for the moments.
This has technical advantages including reducing lattice-spacing, tuning,
and perturbative errors.
In continuum perturbation theory,
\begin{align}
R_4 &= r_{4}(\a_{\MSbar}, \mu) \label{R_4,pert}\\
R_n &= \frac{1}{m_c(\mu)} \, r_{n}(\a_{\MSbar}, \mu) \quad (n \geq 6) \, .
\label{R_n,pert}
\end{align}
Here $r_n$ are the perturbative expressions derived from appropriate
powers of $g_n/g_n^{(0)}$, with $g_n^{(0)}$ the lowest order perturbative
result.
Taking $\mu=3 m_h(\mu)$ in the analysis,
for a given $m_{0h}$ one computes the values of $R_n$
from \EQ{R_n,latt} and gets an estimate of $m_c(3m_h) = R_n/r_n(3m_h)$,
via \EQ{R_n,pert}, and in this way both the value and scale dependence
of $m_c(\mu)$ can be determined.

\begin{figure}
\begin{center}
\includegraphics[width=0.5\textwidth]{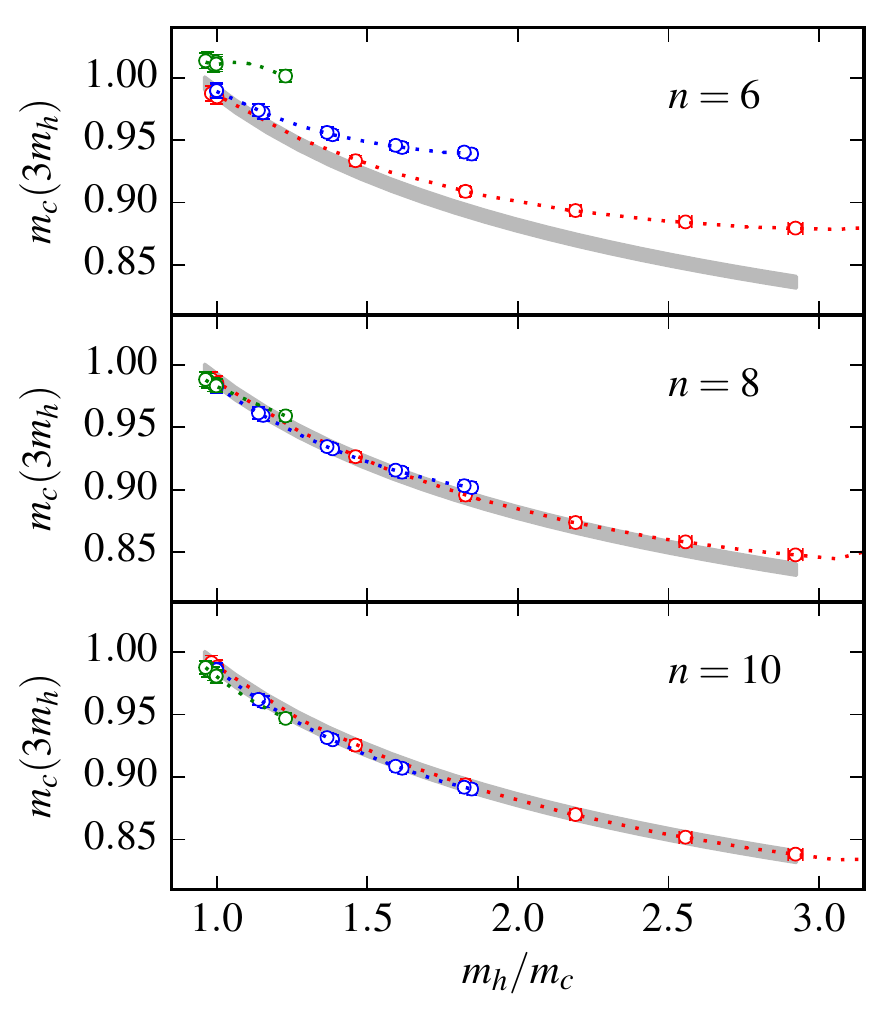}
\end{center}
\caption{Results of a current-current correlator analysis reproduced
from~\cite{Chakraborty:2014aca}.
The colored circles give the lattice data from reduced moments
at successively finer lattice spacings, while the gray band shows the
best-fit value of $m_c$ evolved using perturbation theory. Discretization
artifacts are evident in the data for low moments but by $n=10$ the data
has essentially collapsed onto the perturbative curve.
\label{fig:mc-mh}}
\end{figure}

Results of this procedure from~\cite{Chakraborty:2014aca}
are reproduced in Fig.~\ref{fig:mc-mh}.
One sees that the low moment data has noticeable lattice artifacts,
and that by $n=10$ lattice artifacts are sufficiently small that the
data collapses onto gray curve which gives the perturbative evolution.

\subsection{MRS masses}
This new entrant to the theorist's
toolkit~\cite{Komijani:2017vep,Brambilla:2017hcq}
takes as its departure point
the factorization of heavy meson masses into heavy and light degrees
of freedom based on Heavy Quark Effective Theory (HQET):
\begin{equation} \label{MRS}
M_H = m_Q + \overline{\Lambda} + \frac{\mu_\pi^2}{2m_Q} -
\frac{\mu_G^2(m_Q)}{2m_Q} + \dotsb \,,
\end{equation}
Here $m_Q$ is the
pole mass of the heavy quark $Q$
and $\frac{\mu^2_\pi}{2 m_Q}$ its kinetic energy,
while the next two terms
parameterize light degrees of freedom in the meson, namely
the energy of light quarks and gluons $\overline{\Lambda}$ and
$\frac{\mu_G^2(m_Q)}{2m_Q}$ the hyperfine energy due to heavy quark spin.

The Fermilab Lattice, MILC and TUMQCD 
collaborations~\cite{FermilabLattice:2018est}
map out the heavy meson mass dependence
on lattice input quark mass from $D$ to $B$, using a
fictitious valence mass $m_{h,0}$, this data forms the
main input to their fits. What one desires is however
not how the heavy meson mass depends on the bare input mass,
but 
how the heavy meson mass depends on the $\MSbar$
renormalized mass (or equivalently how the $\MSbar$
quark mass depends on the known meson masses). Evaluating
this function using the precisely known meson masses
$m_{D_s}$, $m_{B_s}$ will then give the charm,
bottom quark mass respectively. This is where
Eq.~\eqref{MRS} comes in.
The challenge then lies in relating the pole mass to
the desired $\MSbar$ mass.

The perturbative series connecting the pole mass
to the $\MSbar$ mass diverges due to
renormalons~\cite{Komijani:2017vep,Brambilla:2017hcq}, but in the context of~\EQ{MRS} the authors
introduce another mass definition, the so-called minimal
renormalon subtracted (MRS) mass, which
removes the leading renormalon ambiguity from the pole mass
and has a well-behaved
connection to the $\MSbar$ scheme.


\newcommand{\OLambda}{\overline{\Lambda}}
\newcommand{\Om}{\overline{m}}
\begin{align*}
m_{\text{pole}} + \OLambda &=
\Om \,
\Biggl( 1+\sum_{n=0}^{\infty} r_n \, \a_s^{n+1}(\Om) \Biggr) +
\OLambda \rightarrow \\ &\overline{m} \,
\Biggl( 1+\sum_{n=0}^{\infty} [r_n - R_n] \, \a_s^{n+1}(\Om) \Biggr)
+ J_{\text{MRS}}(\Om) + \left[ \delta_m + \OLambda \right] \\
&= m_{\text{MRS}} + \OLambda_{\text{MRS}} \,,
\end{align*}
where $J_\text{MRS}$ sums the $R_n\,\alpha_s^{n+1}$ series, and can be computed from a convergent series in $1/\alpha_s$.

The effectiveness of expressing Eq.~\eqref{MRS} in terms of
the MRS mass can be seen from the series coefficients
$r_n$ and $R_n$:
\begin{align*}
r_n &=  (0.4244, 1.0351, 3.6932, 17.4358, \dotsc) \\
R_n &= (0.5350, 1.0691, 3.5966, 17.4195, \dotsc) \\
r_n - R_n &=  (-0.1106, -0.0340, 0.0966, 0.0162, \dotsc) \,.
\end{align*}
Although individually poorly behaved as $n$ increases, the
difference $r_n - R_n$ relating the MRS mass to the pole mass
is well under control.
In this way the fit form can be expressed in terms of
the MRS (or equivalently $\MSbar$) mass.

\begin{figure}
\includegraphics[width=0.5\textwidth]{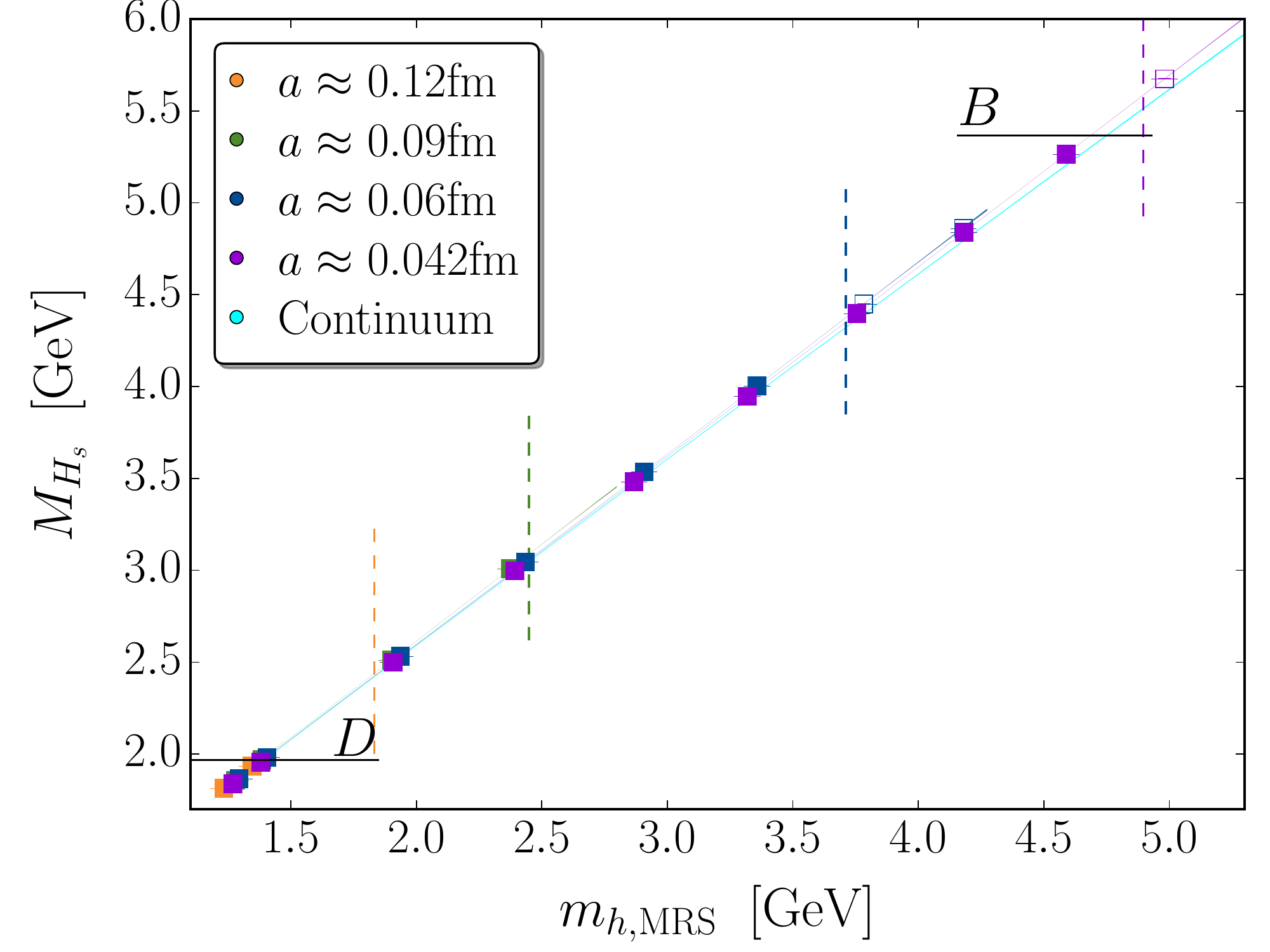}
\includegraphics[width=0.5\textwidth]{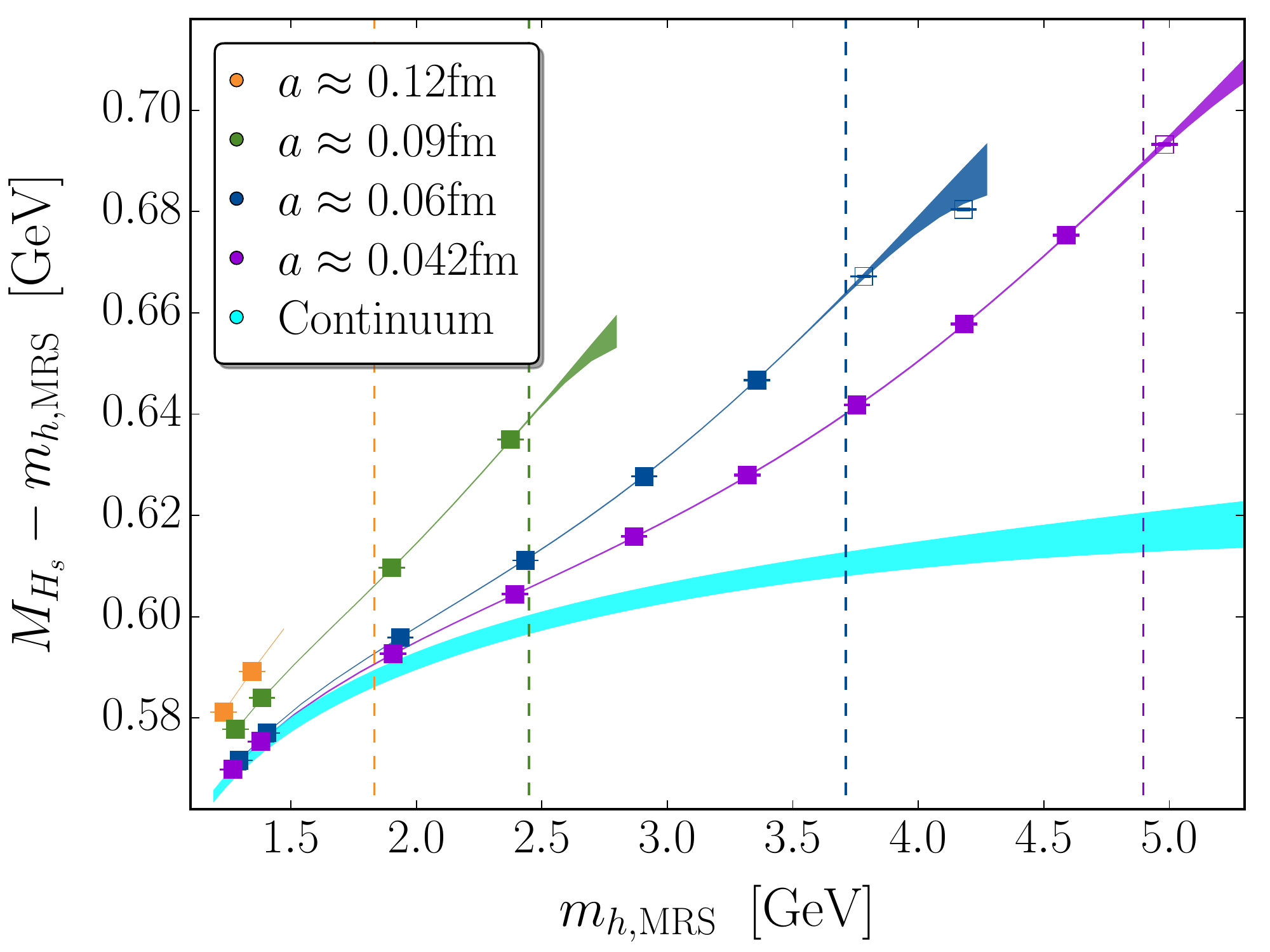}
\caption{(Left) Fit of lattice determined heavy meson masses to
Eq.~\eqref{MRS}, reproduced from~\cite{FermilabLattice:2018est}.
(Right) Data
and fit result
with the leading term $m_{h,\text{MRS}}$ subtracted. From this
one clearly sees the lattice discretization effects, and the
continuum determination of the subleading terms in
Eq.~\eqref{MRS} from the lattice data given by the blue band.
\label{fig:M_vs_mRS}}
\end{figure}

Fitting the data to Eq.~\eqref{MRS} is
shown in Fig.~\ref{fig:M_vs_mRS}. Since the ``signal''
in this figure is just the linear heavy mass dependence,
it is instructive to consider a plot of
$M_{H_s} - m_{h, \text{MRS}}$. Here the lattice artifacts
become clearly visible, and the continuum extrapolation of this
data determines the additional continuum terms in Eq.~\eqref{MRS}.

\subsection{RI/(S)MOM} \label{sec:ri/smom}
While the previous two methods are tailored specifically to quark masses (and $\alpha_s$ for current-current correlators), the
method of non-perturbative renormalization is a very general technique
used for renormalizing operators in lattice field theory~\cite{Martinelli:1994ty}. 
Here we consider
its application to quark masses using staggered fermions~\cite{Lytle:2013qoa}.

The main observation is that off-shell momentum subtraction schemes
can be directly implemented on the lattice (unlike the $\MSbar$ scheme),
provided that the intermediate renormalization scale $\mu$ is well below
the lattice cutoff (otherwise lattice artifacts will be large), and that
a perturbative calculation in the continuum can be used to then convert
to $\MSbar$, provided $\mu$ is in the ``perturbative'' regime.
\begin{equation}
\Lambda_{\text{QCD}} \ll \mu \ll \pi/a
\end{equation}

In practice, the precision application of this technique is aided by
theoretical and technical improvements, such as using a ``symmetric''
subtraction point~\cite{Sturm:2009kb} and twisted boundary conditions~\cite{Arthur:2010ht}, as well as access
to ensembles at many different lattice spacings, which in this case
are made available by the MILC collaboration.
In concert these
improvements allow one to effectively widen the window condition and
separately disentangle both discretization artifacts and non-perturbative
contributions. 

RI/SMOM schemes were used with staggered fermions in \cite{Lytle:2018evc} to achieve percent-level precision for $m_c$.
The main result is illustrated in \FIG{mc_smom}.
The computation was carried out using three lattice spacings down
to $a \approx 0.06$ fm.
The different
colored lines in the figure correspond to different choices for the
intermediate renormalization scale $\mu$. The
increasing slope of these lines is due to $(a \mu)^2$ lattice artifacts,
and the curvature at larger $\mu$ shows presence of $a^4$ effects.
That these these lines do not converge to a single point at $a=0$ indicates
the presence of non-perturbative effects, which are suppressed at increasing
$\mu$ values. The final result for $m_c(3 \GeV)$ is given by the gray dot
on the left and compared with the current-current correlator result.

\subsubsection{Perturbative matching}
\begin{figure}
\begin{center}
\includegraphics[width=0.7\textwidth]{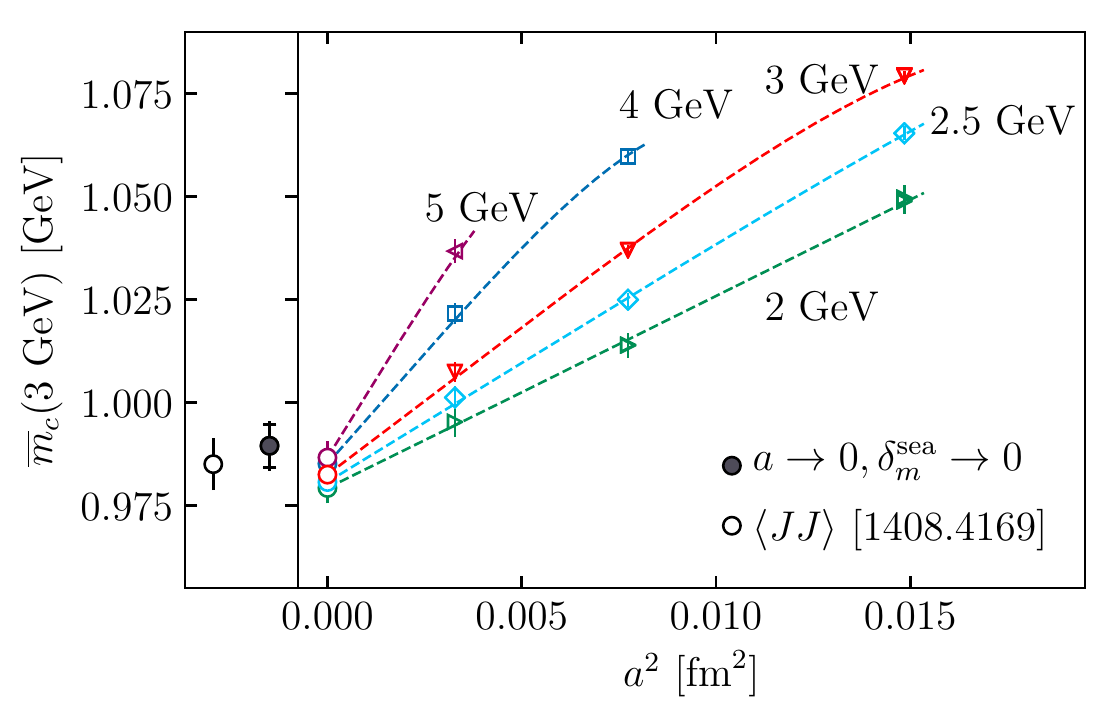}
\end{center}
\caption{Determination of $m_c^{\MSbar}(3 \GeV)$ based on the RI/SMOM
intermediate scheme using highly improved staggered fermions, reproduced
from~\cite{Lytle:2018evc}. The calculation was carried out at three
lattice spacings and a range of intermediate scales $\mu \in [2,5]$ GeV
to quantify discretization and condensate effects. The final result is
given by the gray point on the left and compared with the result based on current-current correlators~\cite{Chakraborty:2014aca}.
\label{fig:mc_smom}}
\end{figure}

Perturbative matching factors, computed in the continuum, are a key ingredient
for these determinations, used to relate the lattice-determined mass
renormalization constant to the conventional $\MSbar$ scheme.
\begin{equation}
    m_c^{\MSbar} = C_m^{\MSbar/ \text{SMOM}}(\mu) 
    Z_m^{\text{SMOM}}(\mu) m_{c,0}
\end{equation}
In 2018 the SMOM $\to \MSbar$ conversion factor,
known at $\O(\alpha_s^2)$ \cite{Gorbahn:2010bf,Almeida:2010ns},
was a leading source of uncertainty with the unknown $c_\alpha \alpha_s^3$
term estimated at $0.22 \%$ from the fit shown in 
Fig.~\ref{fig:mc_smom}. This term was subsequently
calculated by two sets of authors in~\cite{Kniehl:2020sgo} and \cite{Bednyakov:2020ugu},
\begin{equation}
C_m^{\MSbar/\text{SMOM}}(n_f=4, 3 \text{ GeV}) = 
1 - 0.01307 - 0.00269 - \green{0.00196}
\end{equation}
and an update of the result estimates the new $c_\alpha \alpha_s^4$ at
$\approx 0.1\%$.

\subsubsection{Adding QED}
To really go beyond this level of precision,
one must contend with the effects
of QED, which could naively be as large as $1 \%$.
The effects of (quenched) QED were quantified in the RI/SMOM framework
by extending~\cite{Lytle:2018evc} 
in~\cite{Hatton:2020qhk} and~\cite{Hatton:2021syc} 
for $m_c$ and $m_b$ respectively.
In this approximation the QED field is simply multiplied into the
SU(3) gauge field.
The effect of the quenching approximation neglects terms of size
$\alpha_s^2 \alpha$, and so is expected to be maybe $10 \%$ of the overall
tiny correction.
The effect of strong isospin breaking is also neglected, which for
the quantities considered 
is expected to be sub-0.1\%~\cite{Hatton:2020qhk}.
The overall effect of electromagnetism
decreases $m_c$ by $0.18(2) \%$, while the ratio $m_b/m_c$ was found to
increase by $0.17(3) \%$, confirming that the neglect of these corrections
is permissible at the percent-level, but becomes significant beyond that.

\section{Bottom quark mass} \label{sec:bottom}
\begin{figure}
\begin{center}
\includegraphics{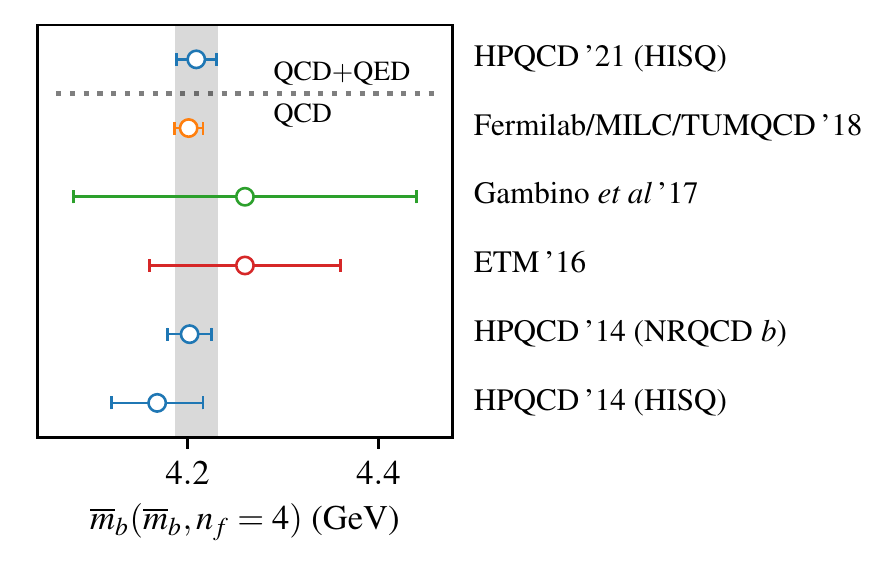}
\end{center}
\caption{Summary plot of $n_f=4$ $m_b^{\MSbar}(m_b)$ results from lattice QCD, reproduced from~\cite{Hatton:2021syc}. The result HPQCD 21, published since the 2019 FLAG review, includes the effects of (quenched) QED.
\label{fig:mbm}}
\end{figure}

There are 5 new lattice determinations of the bottom mass since 2015
 (bringing the total to 9), 
and one since the 2019 FLAG review. The FLAG review gives an average
\begin{equation}
m_b^{\MSbar}(m_b) = 4.198(12) \qquad (n_f = \text{2+1+1}) \,,
\end{equation}
corresponding to a 0.3\% uncertainty!
The results entering this average are shown in \FIG{mbm}, plus the
new result published since the FLAG review from HPQCD~\cite{Hatton:2021syc}. 
The HPQCD result comes from
a determination of the mass ratio $m_b/m_c$, including the 
effects of (quenched) QED, 
multiplied by the determination of $m_c$
via RI/SMOM described in \SEC{ri/smom}. 
It is in agreement with the FLAG average
(as well as with the earlier HPQCD result without QED effects), showing
that QED effects are tiny for this quantity.

\section{Summary \& Conclusion} \label{sec:Summary}
Since Charm 2015 there has been significant progress made in the computation
of heavy quark masses from lattice. For one thing, there are many
new results -- I count 13 since 2015, and 4 since the most recent
FLAG review. These come from several different collaborations using
different lattice discretizations, which gives confidence that
lattice artifacts are under control. But perhaps more importantly,
there are now a variety of different techniques being used, and
the agreement among methods gives confidence that sources of
systematic uncertainty particular to each method are being
well-estimated. 
This establishes the masses at the (sub-)percent level. 
At this level of precision one may expect QED effects to become
relevant. 
These were quantified in~\cite{Hatton:2020qhk} and~\cite{Hatton:2021syc}, and found to be small
($\lesssim 0.2\%$ for charm).

I've focused here on a few of the recent calculations quoting
small uncertainties, emphasizing the differing methodologies used.
However, the good agreement between results
is not without caveats. The main one being the
three calculations highlighted here are based on simulations with HISQ fermions, and use the same sets of FNAL-MILC ensembles.
It will be interesting in the future to see how results using
different regularizations compare.
The methods highlighted here are also 
not the only ones on the market. In particular small volume step-scaling techniques provide a 
promising avenue to precision quark 
masses and are continuing to evolve. Surely there are
additional variants, as well as brand new ideas, waiting to be discovered.

\section{Acknowledgements}
I would like to express my gratitude to the organizers for an enjoyable conference and their hard work ensuring that the conference could go forward. I would also like to thank
Andreas Kronfeld for providing helpful comments on the manuscript,
and acknowledge support by the U.S. Department of Energy under grant
number DE-SC0015655.


\end{document}